\documentclass[aps,prc,twocolumn]{revtex4-2}
\usepackage{amsmath,amssymb}
\usepackage{graphicx}

\newcommand{\be}{\begin{equation}}
\newcommand{\ee}{\end{equation}}
\newcommand{\ba}{\begin{eqnarray}}
\newcommand{\ea}{\end{eqnarray}}
\newcommand{\ban}{\begin{eqnarray*}}
\newcommand{\ean}{\end{eqnarray*}}
\newcommand \nn {\nonumber}

\begin{document}

\title{$\boldsymbol{\phi}$-$\boldsymbol{p}$ bound state and completeness of quantum states}

\author{Arkadiusz Kuro\'s$^1$, Rados\l aw Maj$^1$ and Stanis\l aw Mr\' owczy\' nski$^{1,2}$}

\affiliation{$^1$Institute of Physics, Jan Kochanowski University, ul. Uniwersytecka 7, 25-406 Kielce, Poland 
\\
$^2$National Centre for Nuclear Research, ul. Pasteura 7, 02-093 Warsaw, Poland}

\date{August 21, 2024}

\begin{abstract}

The existence of a bound state of a $\phi$ meson and proton has been recently indicated on the basis of a combined analysis of the measured  $\phi$-$p$ correlation function and the QCD lattice calculations of $\phi$-$p$ interaction in the spin 1/2 and 3/2 channels. We check whether the correlation functions and bound state formation rate satisfy the sum rule which follows from the completeness of quantum states. The sum rule is not satisfied but it is close to it, suggesting that the inferred scattering parameters, which determine the correlation functions, are not fully consistent with the potential responsible for the bound state existence. 

\end{abstract}

\maketitle

\section{Introduction}

A possible existence of a bound state of a $\phi$ meson and proton $p$ has been debated over two decades since it was first considered in \cite{Gao:2000az}. Experimental evidence of the $\phi$-$p$ bound state is lacking and there are theoretical arguments for and against the occurrence of this state. Since the quark content of $\phi$ is mostly $s\bar{s}$ the elastic interaction via exchange of $u$ or $d$ quarks is suppressed due to the Okubo-Zweig-Iizuki (OZI) rule. Therefore, the inelastic channel $\phi p \leftrightarrow K \Lambda$ seems to dominate the $\phi$-$p$ interaction which disfavors the $\phi$-$p$ bound state. However, the existence of  QCD van der Waals forces driven by multi-gluon exchange was argued  in such systems \cite{Brodsky:1989jd}. The forces can generate an attraction sufficient to support the $\phi$-$p$ bound state \cite{Gao:2000az,Huang:2005gw,Gao:2017hya,Sun:2022cxf}. A convincing evidence of the $\phi$-$p$ bound state would be of a great interest. It would improve our understanding of strong interactions, in particular, shed light on multi-quark states different than baryons and mesons as the $\phi$-$p$ bound state can be viewed as a hidden strange pentaquark.

Recently, the ALICE collaboration has managed to measure the femtoscopic correlation function of $\phi$ and $p$ with small relative momenta produced in proton-proton collisions at LHC \cite{ALICE:2021cpv}. The correlation function clearly indicates the attractive $\phi$-$p$ interaction. The inferred scattering length is sizable and, rather unexpectedly, purely real which shows that the interaction is mostly elastic. The measurement, however, does not allow to draw a far going conclusion, as the correlation function and scattering parameters are spin averaged. One should remember that the $\phi$-$p$ system can exist in two spin states: 1/2 and 3/2. 

An important step to make full use of the unique experimental data \cite{ALICE:2021cpv} has been taken in the very recent work \cite{Chizzali:2022pjd}. Combining the measured correlation function and lattice calculations of $\phi$-$p$ interaction by HAL QCD collaboration \cite{Lyu:2022imf}, the authors managed to obtain parametrizations of interaction potential and scattering parameters for each spin channel. The results imply the existence of $\phi$-$p$ bound state with the spin-parity $J^P = {\frac{1}{2}}^-$ and binding energy $E_B = -(12.8 \div 56.1)$~MeV \cite{Chizzali:2022pjd}. 

The aim of this paper is to check whether the correlation functions of spin 1/2 and 3/2 channels and the bound state formation rate, which directly follow from the results of the study \cite{Chizzali:2022pjd}, satisfy the sum rule \cite{Mrowczynski:1994rn,Maj:2004tb,Maj:2019hyy} resulting from the completeness of quantum states of the $\phi$-$p$ system. 

The femtoscopic correlation function integrated over particle relative momentum was shown \cite{Mrowczynski:1994rn} to obey a sum rule due to the completeness of quantum states. In case of identical particles, {\it e.g.} $\pi^\pm \pi^\pm$, the integrated correlation function equals a particle's density in their source. In case of nonidentical particles as neutron and proton, the integral vanishes in the spin 0 channel and it is related to a deuteron formation rate in the spin 1 channel. The sum rule shows that the correlation can be negative with the correlation function smaller than unity even so the inter-particle interaction is attractive. It happens when the interaction is absorptive as in the case of spin 1 channel of neutron-proton interaction where the deuteron formation is possible \cite{Mrowczynski:1992gc}. Such a situation is observed in the  experimentally studied proton-antiproton correlations \cite{ALICE:2019igo} because of the annihilation process. We also note that the $\phi$-$p$ correlation function of the spin 1/2 channel extracted in \cite{Chizzali:2022pjd} is smaller than unity which immediately suggests the existence of $\phi$-$p$ bound state in this channel. 

In principle, the sum rule requires the integration of the correlation function up to infinite momentum. However, one expects that the integral saturates at a sufficiently big momentum. Indeed, in case of free identical particles, when the correlation is solely due to quantum statistics, the sum rule is satisfied when the integration extends up to the momentum $q_{\rm max}$ which equals a few times the inverse radius of particle source \cite{Maj:2004tb}. When the sum rule is applied to correlation functions of interacting particles the momentum integral appears to be divergent, as the correlation functions tend to unity not sufficiently fast as the momentum grows \cite{Maj:2004tb}. 

The problem is resolved \cite{Maj:2019hyy} by considering the sum rule not of a single correlation function but of the difference or sum of two appropriately chosen correlation functions. The ultraviolet divergence is then canceled out but the physical content of the sum rule remains unchanged. For example, the sum rule tells us that the integral of the neutron-proton correlation function of the spin 0 channel vanishes while that of the neutron-proton correlation function of the spin 1 channel is related to the deuteron formation rate. So, the integral over the difference of the correlation functions is again related to the deuteron formation rate but the integral is convergent. One can say that the spin 0 correlation function plays a role of the regulator of the integral over the spin 1 correlation function. It was checked \cite{Maj:2019hyy} that the procedure works perfectly well for the exact Coulomb and neutron-proton correlation functions. In this paper we apply the sum rule to the $\phi$-$p$ correlation function in the spin 1/2 channel, treating the spin 3/2 correlation function as the regulator. 

\section{Sum rule}
\label{sec-sum-rule}

As discussed in detail in \cite{Maj:2019hyy}, the sum rule in case of $\phi$ and $p$ can be written as
\be 
\label{sum-rule}
\int d^3 q \, \Big( C_{1/2} ({\bf q}) - C_{3/2} ({\bf q}) \Big)
=  -  A , 
\ee
where $C_{1/2} ({\bf q})$ and $C_{3/2} ({\bf q})$ are the $\phi$-$p$ correlation functions in the spin 1/2 and 3/2 channels and ${\bf q}$ is the relative momentum. The spin averaged correlation function, which was studied experimentally \cite{ALICE:2021cpv}, equals
\be
C ({\bf q}) \equiv \frac{1}{3}C_{1/2} ({\bf q}) + \frac{2}{3} C_{3/2} ({\bf q}) .
\ee
The weight factors 1/3 and 2/3 reflect numbers of spin states in the two channels under the assumption that $\phi$ and $p$  are unpolarized. The spin averaged correlation function is defined in the following way 
\be
\label{corr-function-def}
\frac{dP_{\phi p}}{d^3p_\phi  d^3p_p} = C({\bf q}) \, 
\frac{dP_\phi}{d^3p_\phi}  \frac{dP_p}{d^3p_p} ,
\ee
where $\frac{dP_\phi}{d^3p_\phi}$,  $\frac{dP_p}{d^3p_p}$ and $\frac{dP_{\phi p}}{d^3p_\phi  d^3p_p}$ are probability densities to observe a $\phi$ meson with momentum ${\bf p}_{\phi}$, a proton with ${\bf p}_p$ and a pair $\phi p$ with momenta ${\bf p}_{\phi}$ and ${\bf p}_p$ in a collision final state. The quantity $A$, which enters Eq.~(\ref{sum-rule}) is the formation rate of the $\phi$-$p$ bound state in the spin 1/2 channel. We assume here that there is only one bound state in the spin 1/2 channel and no bound states in the 3/2 channel. The formation rate is defined analogously to the correlation function (\ref{corr-function-def}) that is
\be
\label{form-rate-def}
\frac{dP_{(\phi p)}}{d^3p_{(\phi p)}} =\frac{1}{3} A \,
\frac{dP_\phi}{d^3p_\phi}  \frac{dP_p}{d^3p_p} \bigg|_{{\bf q}=0},
\ee
where $\frac{dP_{(\phi p)}}{d^3p_{(\phi p)}}$ is a probability density to observe the $\phi$-$p$ bound state with momentum ${\bf p}_{(\phi p)}$. The factor 1/3 reflects the numbers of spin states in the spin 1/2 channel. The momenta ${\bf p}_{(\phi p)}$, ${\bf p}_{\phi}$ and ${\bf p}_p$ are such that ${\bf p}_{(\phi p)} = {\bf p}_{\phi} + {\bf p}_p$ and the relative momentum ${\bf q}$ vanishes in the center-of-mass frame of the $\phi$-$p$ system. The binding energy of the bound state is assumed to be much smaller than its mass. 

The sum rule (\ref{sum-rule}) does not take into account the inelastic transitions like $\phi p \leftrightarrow K^+ \Lambda$ that is the states of $K^+$ and $\Lambda$ are not included in the complete set of states of $\phi$ and $p$. We return to this important problem in Sec.~\ref{sec-corr-fun}. 

The correlation functions $C_s({\bf q})$ with $s=1/2$ and $s=3/2$ can be expressed as \cite{Koonin:1977fh} 
\be
\label{fun-corr-relative}
C_s({\bf q}) = \int d^3 r \, S_r ({\bf r}) |\phi^s_{\bf q}({\bf r})|^2 ,
\ee
where $S_r ({\bf r})$ is the source function which is the normalized to unity distribution of relative distance of $\phi$ and $p$ at the moment of emission from the source and $\phi^s_{\bf q}({\bf r})$ is the $\phi$-$p$ wave function in the scattering state of spin $s$ and momentum ${\bf q}$. The formation rate is expressed in a similar way that is \cite{Sato:1981ez} 
\be 
\label{form-rate}
A =  (2\pi)^3 \int d^3r \, S_r({\bf r}) \vert \phi_B({\bf r}) \vert^2 ,
\ee
where $\phi_B({\bf r})$ is the wave function of the $\phi$-$p$ bound state. The source function is usually chosen in the Gaussian form 
\be 
\label{Gauss-source-r}
S_r({\bf r}) = {1 \over (4 \pi r_0^2)^{3/2}} \,
{\rm exp} \Big(-{{\bf r}^2 \over 4r_0^2}\Big) .
\ee 
which gives the mean radius squared of a single-particle source equal $\langle {\bf r}^2 \rangle = 3 r_0^2$.

Two comments are in order here. The formulas (\ref{fun-corr-relative}) and (\ref{form-rate}) are written in a non-relativistic form even though the particles of interest are typically relativistic. The correlation function, however, significantly differs from unity only for small relative velocities of $\phi$ and $p$. Similarly, the formation rate is sizable when the relative velocity is small. Therefore, in the center-of-mass frame the relative motion of $\phi$ and $p$ can be treated as non-relativistic and the corresponding wave functions are solutions of the Schr\"odinger equation. 

We also note that the formulas (\ref{fun-corr-relative}) and (\ref{form-rate}) are written as for the instantaneous emission of $\phi$ and $p$. However, a time duration of the emission process can be easily incorporated \cite{Koonin:1977fh}. If one uses the isotropic Gaussian source function (\ref{Gauss-source-r}), the time duration $\tau$ simply enlarges the effective radius of the source from $r_0$ to $\sqrt{r_0^2 + v^2\tau^2}$ where $v$ is the velocity of the particle pair relative to the source. Then, the source function $S_r ({\bf r})$ effectively gives the space-time separation of the emitted particles. 

\section{Is the sum rule satisfied?}
\label{sec-sum-rule-check}

To check whether the sum rule (\ref{sum-rule}) is satisfied by the correlation functions and bound state formation computed with the parameters of $\phi$-$p$ system inferred in the study \cite{Chizzali:2022pjd} one needs the wave functions of scattering and bound states to compute the correlation functions (\ref{fun-corr-relative}) and formation rate (\ref{form-rate}).

\begin{figure}[t]
\centering
\includegraphics[width=8cm]{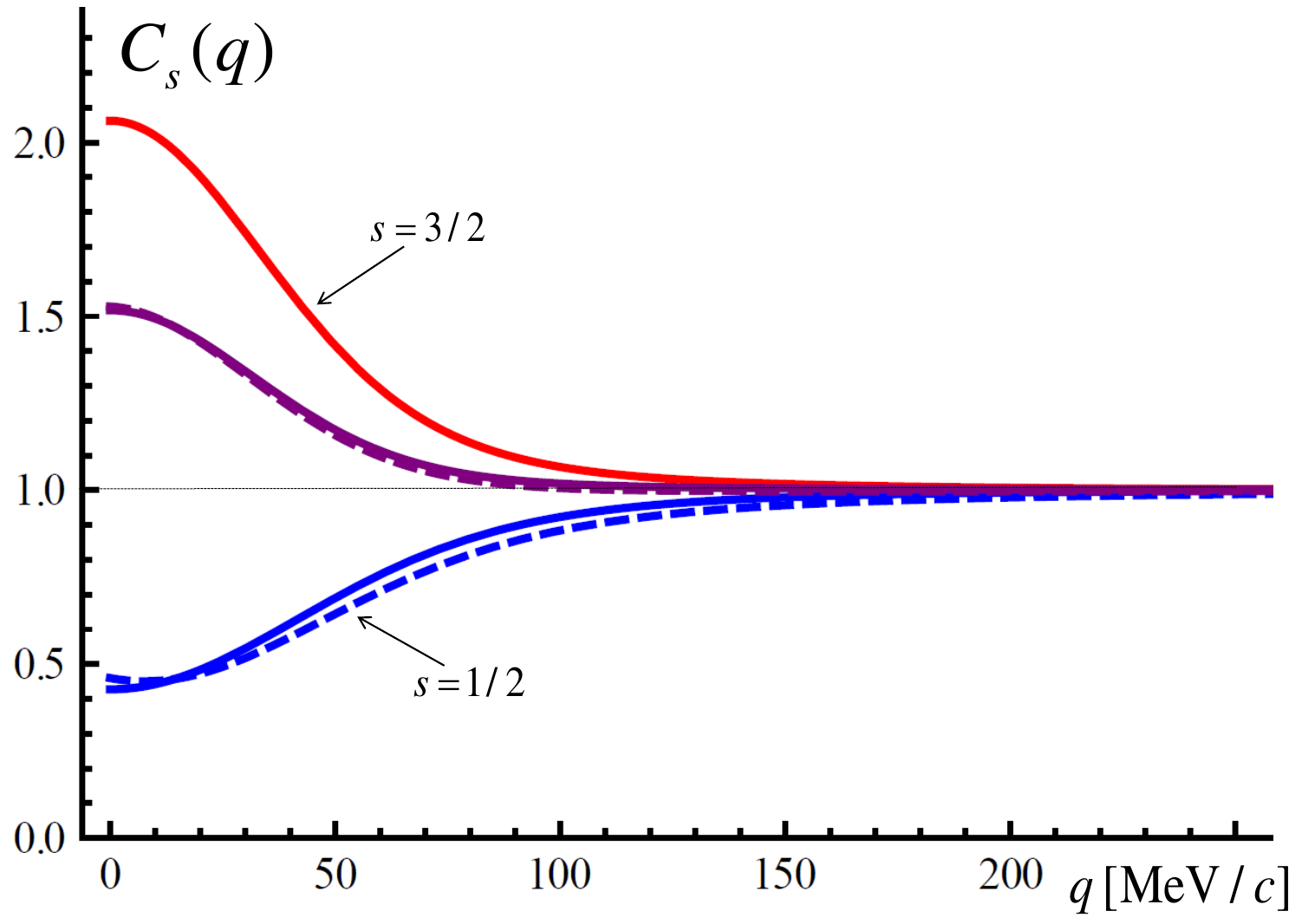}
\vspace{-3mm}
\caption{The doublet (blue line), quartet (red line) and spin average (magenta line) correlation functions for $r_0 = 2$~fm. The solid lines correspond to the face values of the scattering parameters while the dashed lines are computed with the maximal imaginary parts of the parameters.}
\label{fig-corr-functions}
\end{figure}

\subsection{Correlation functions}
\label{sec-corr-fun}
If the source radius is significantly bigger than the interaction range, the correlation functions (\ref{fun-corr-relative}) can be computed as proposed in \cite{Lednicky:1981su} that is using the scattering wave function in the asymptotic form 
\be 
\label{scatter}
\phi_{\bf q}^s({\bf r}) = e^{i{\bf q}{\bf r}}  + f_s(q) \, \frac{e^{iq r}}{r} , 
\ee
with the $s$-wave scattering amplitude
\be 
\label{ampli}
f_s (q)  = \frac{- a_s}{1 - \frac{1}{2} d_s a_s q^2 + i q a_s} , 
\ee 
where $a_s$ and $d_s$ are the scattering lengths and effective ranges in the spin channels $s=1/2$ and $s=3/2$.  In Table~\ref{table-scatt-parameters} we give values of the scattering parameters obtained in the study \cite{Chizzali:2022pjd} with the maximal error which is the sum of statistical and systematical errors. 

\begin{table}[b]
\centering 
\begin{tabular}{ c |c | c | c | c } 
\hline
& $\Re a$ [fm] & $\Im a$ [fm] & $\Re d$ [fm]  & $\Im d$ [fm] \\
\hline  
\, $s=\frac{1}{2}$ \, & \, $1.54^{+0.62}_{-0.69}$ \, & \, $0.00^{+0.00}_{-0.51}$ \, &
\, $0.39^{+0.11}_{-0.12}$ \, & \, $ 0.00^{+0.00}_{-0.06} $ \,  \\ [2mm]
\,$s=\frac{3}{2}$ \, & \, $ -1.43^{+0.59}_{-0.29}$ \, & 0 & \,$2.36^{+0.12}_{-0.58}$ & 0    \\
\hline
\end{tabular}
\caption{The scattering lengths and effective ranges in spin 1/2 and 3/2 channels.}
\label{table-scatt-parameters}
\end{table}

The doublet ($s=1/2$), quartet ($s=3/2$) and spin average correlation functions computed using the wave functions (\ref{scatter}) with the parameters from Table~\ref{table-scatt-parameters} and the Gaussian source with $r_0 = 2$~fm are shown in Fig.~\ref{fig-corr-functions}. The solid lines correspond to the face values of the scattering parameters that is $a_{1/2} = 1.54$ fm,  $d_{1/2} = 0.39$ fm and $a_{3/2} = - 1.43$ fm,  $d_{3/2} = 2.36$ fm while the dashed lines are computed with the maximal imaginary parts of the parameters that is $a_{1/2} = 1.54 - 0.69i$ fm,  $d_{1/2} = 0.39-0.06i$ fm and $a_{3/2} = - 1.43$ fm,  $d_{3/2} = 2.36$ fm. 

As one sees the doublet correlation function is smaller than unity and thus it represents the negative correlation even so the interaction in the $s=1/2$ channel is attractive, see Sec.~\ref{sec-bound-state}. The sum rule \cite{Mrowczynski:1994rn,Maj:2004tb,Maj:2019hyy} clearly shows that such a behavior indicates the existence of a bound state, as already mentioned in the introduction. 

\begin{table*}[t]
\centering 
\begin{tabular}{ c |c | c | c | c | c | c | c} 
\hline
$\beta$ & $a_1$ [MeV] & $b_1$ [fm] & $a_2$ [MeV]  & $b_2$ [fm] & $\gamma$ &
~ $ a_3 m_\pi^4 ~ \rm{[MeV \, fm^2]}$ ~ & $b_3$ [fm] \\
\hline  
~ $6.9^{+1.1}_{-0.6}$ ~ & ~ $-392 \pm 10$ ~ & ~ $0.128 \pm 0.003$ ~ 
& ~ $ - 145 \pm 9$ ~  & ~ $0.284 \pm 0.007$ ~ & ~ $0.0^{+0.0}_{-5.4}$ ~ & 
~ $83 \pm 1$ ~ & ~ $0.582 \pm 0.006$ ~ \\ 
\hline
\end{tabular}
\caption{Parameters of the $\phi$-$p$ interaction potential in the spin 1/2 channel.}
\label{table-potential}
\end{table*}

One also sees in Fig.~\ref{fig-corr-functions} that taking into account the maximal imaginary contributions to the scattering parameters changes the doublet and spin average correlation functions very little. (The scattering parameters of the quartet channel are pure real.) It agrees with the somewhat unexpected conclusion of the study \cite{Chizzali:2022pjd} that the $\phi$-$p$ interaction is strongly dominated by the elastic channel. However, this is not the whole story about a role of imaginary parts of the scattering parameters. 

\begin{figure}[t]
\centering
\includegraphics[width=8.5cm]{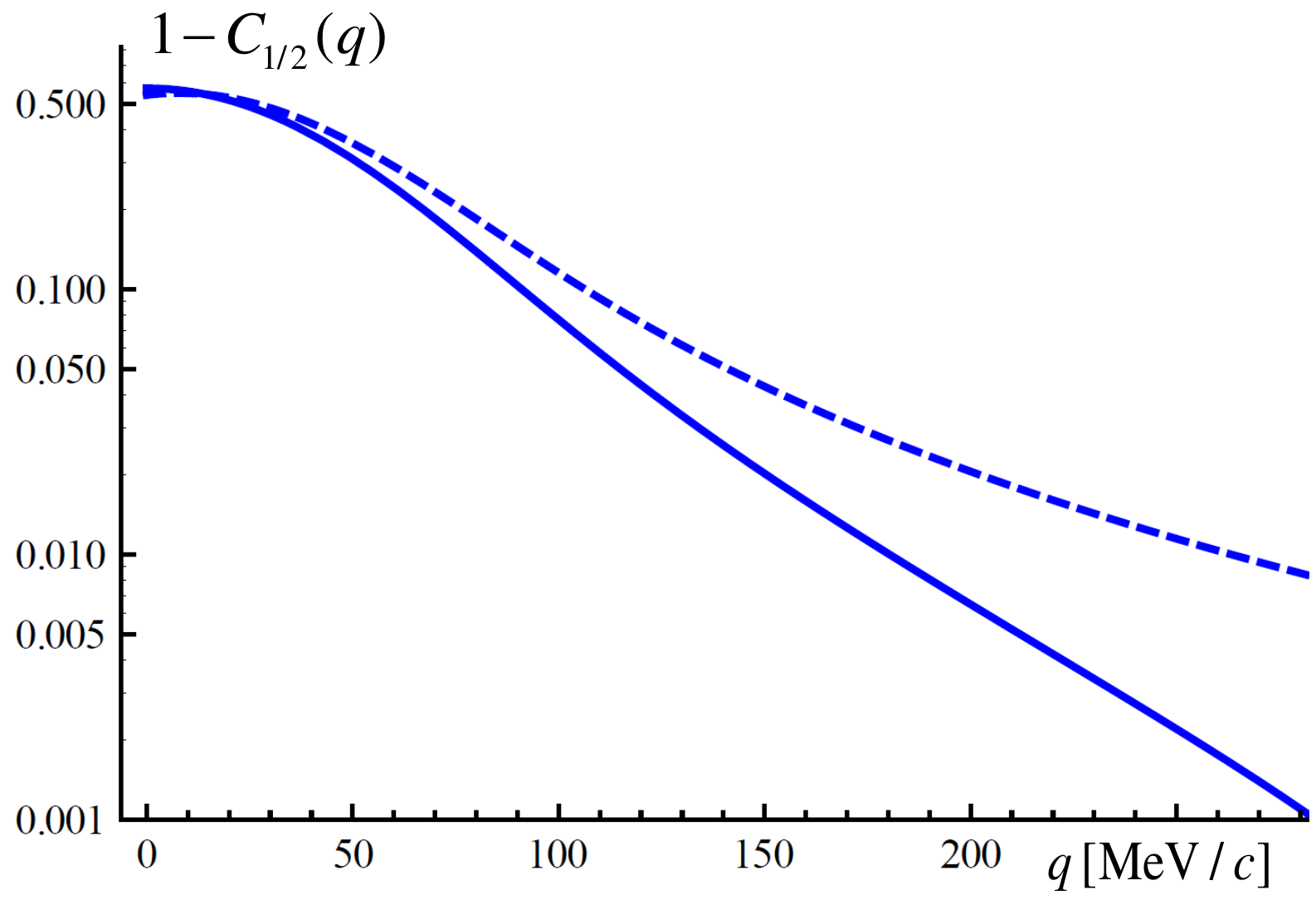}
\vspace{-3mm}
\caption{The doublet correlation functions for $r_0 = 2$~fm. The solid line corresponds to the face values of the scattering parameters while the dashed line is computed with the maximal imaginary parts of the parameters.}
\label{fig-corr-asymptotics}
\end{figure}

It occurs that even small imaginary contributions to the scattering parameters strongly influence asymptotics of the correlation function. In Fig.~\ref{fig-corr-asymptotics} we show the function $1 - C_{1/2}(q)$ computed for $r_0 = 2$ fm in the logarithmic scale. The solid line corresponds to the face values of the scattering parameters, which are pure real, while the dashed line takes into account the maximal imaginary contributions to the scattering length $a_{1/2}$ and effective range $d_{1/2}$. As one can see, the correlation function with purely real parameters tends to unity faster than the one with complex parameters.

We have already mentioned in the introduction that the integral $\int d^3q \big(C_{1/2}(q) - 1 \big)$ is divergent and it needs a regularization.  As argued in \cite{Maj:2019hyy}, the correlation function $C_{3/2}(q)$ can be used as a regulator and indeed the integral $\int d^3q \big(C_{1/2}(q) - C_{3/2}(q)\big)$ is convergent provided the scattering parameters of the function $C_{1/2}(q)$ are real, as are real the parameters of the function $C_{3/2}(q)$. 

We have not found a method to properly regularize the correlation function computed with the complex scattering parameters.  It presumably requires a modification of the sum rule by explicitly including inelastic channels which in turn demands taking into account quantum states that are not mutually orthogonal. The whole problem becomes rather difficult while our analysis from Sec.~\ref{sec-bound-state} shows that an influence of the imaginary part of the potential on the $\phi$-$p$ bound state is very small, suggesting again minor role of inelastic channels. So, we limit our analysis to purely elastic $\phi$-$p$ interaction in accordance with the conclusion of the study \cite{Chizzali:2022pjd}.

\subsection{Bound-state formation rate}
\label{sec-bound-state}

To compute the formation rate (\ref{form-rate}), which enters the sum rule (\ref{sum-rule}), one needs the $s$-wave bound state wave function of the form
\be
\phi_B({\bf r}) = \frac{u(r)}{\sqrt{4 \pi} \, r} .
\ee
The radial wave function $u(r)$ is normalized as
\be
\int_0^\infty dr |u(r)|^2 =1
\ee
and it satisfies the Schr\"odinger equation
\be
\label{Schroedinger-eq}
\Big(- \frac{1}{2 \mu} \frac{d^2}{d r^2}  +  V_{1/2}(r) \Big) u(r) = E_B u(r) ,
\ee
where $\mu = 488.6$~MeV is the reduced mass of $\phi$ and $p$ and $E_B$ is the binding energy. The $\phi$-$p$ interaction potential in the spin 1/2 channel as given in the study \cite{Chizzali:2022pjd} is
\ba
\nn
 V_{1/2}(r) &=& \beta \sum_{i=1,2}a_i e^{-(r/b_i)^2} 
 + a_3 \, m_\pi^4 f(r) \, \frac{e^{-2m_\pi r}}{r^2} 
 \\ [2mm] \label{potential-1/2}
&& ~~~~~~~~~~~~~~~~~~~~~~\,
+ \,i \gamma  f(r) \, \frac{e^{-2m_K r}}{m_K r^2}
\ea
with the form-factor $f(r)$ 
\be
f(r) \equiv \big(1 -  e^{-(r/b_3)^2} \big)^2 .
\ee
The values of parameters of the potential are collected in Table~\ref{table-potential} where we give again the maximal error which is the sum of statistical and systematical errors. The first term in Eq.~(\ref{potential-1/2}) represents the short range repulsion, the second one the long range attraction due to two pion exchange and the third imaginary term represents the inelastic interaction due to two kaon exchange. The imaginary part of the potential is numerically much smaller than the real part for any $r$. 

\begin{figure}[b]
\centering
\includegraphics[width=7.5cm]{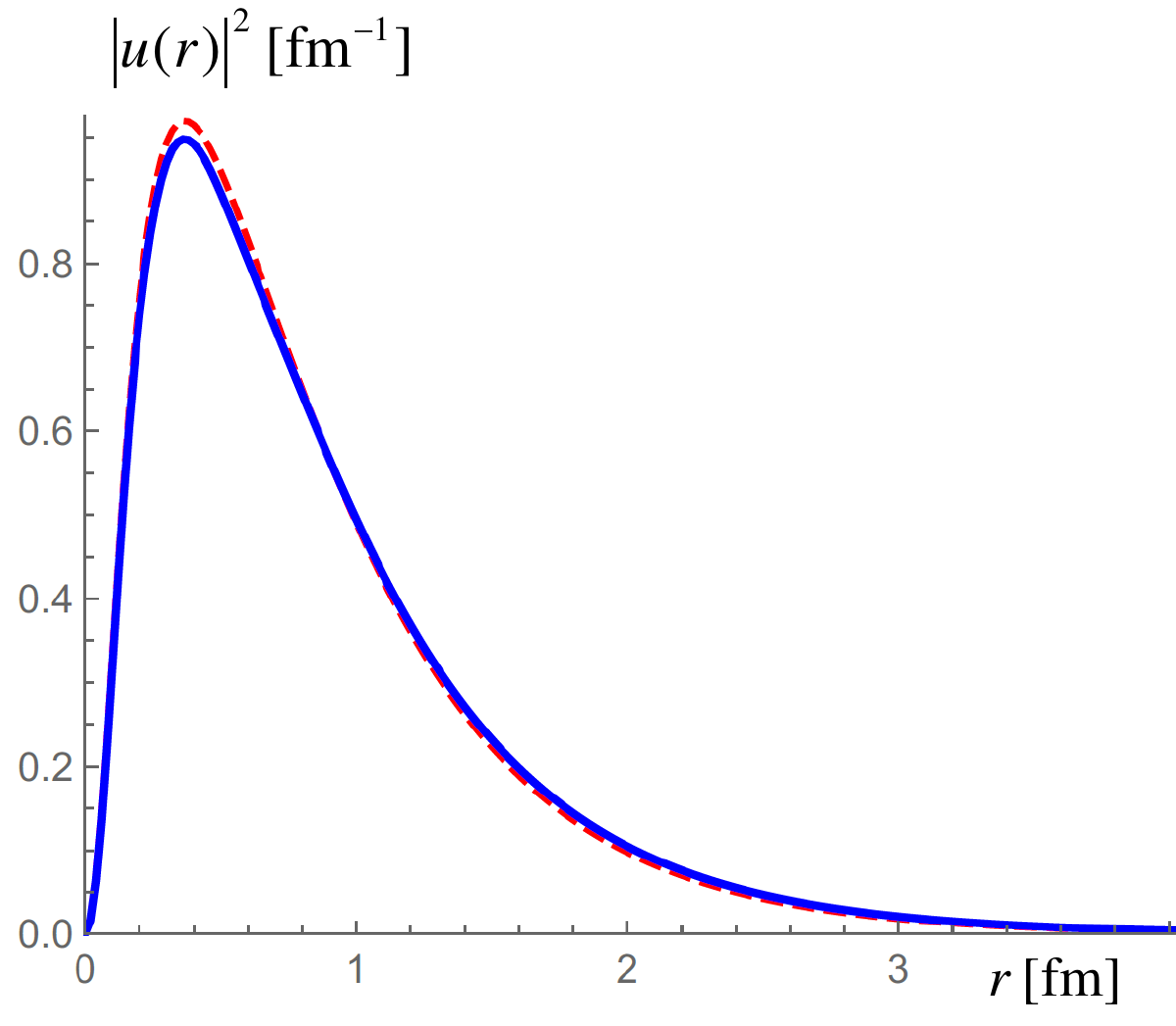}
\vspace{-3mm}
\caption{The modulus square of the bound state wave function computed with the face values of potential parameters (solid blue line) and with the face values of the potential parameters except $\gamma = - 5.4$ (dashed red line).}
\label{fig-wave-fun-potential}
\end{figure}

We have used the optimized Rayleigh-Ritz variational method to solve the time-independent Schr\"odinger equation (\ref{Schroedinger-eq}) with the potential (\ref{potential-1/2}). The wave function $u(r)$ is expanded into a finite set of basis functions with a single nonlinear variational parameter $\mu$ 
\be 
\label{wave-function}
u(r)=\sum_{n=0}^{N-1} c_n \varphi^\mu_n(r) .
\ee
The coefficients $c_n$ and the parameter $\mu$, on which the basis functions depend non-linearly, are determined by the variational principle. The method provides upper bounds of eigenvalues that approach the exact eigenenergies monotonically from above on increasing the number $N$ of basis functions. An accuracy of the wave function can be systematically improved by increasing the basis size and it can be tested using the measure
\be
\eta = \langle u | [\hat{\Lambda}, \hat{H}] | u \rangle,
\ee
which, according to the hypervirial theorem \cite{Epstein:1961zz}, vanishes for any operator $\hat\Lambda$ if $u(r)$ is an eigenfunction of the Hamiltonian $\hat{H}$.

We have used two rather different basis of trial functions. The first one is of trigonometric functions
\be
\label{trigonometric-basis}
\varphi_n^L(r) = \sqrt{\frac{2}{L}} \, \sin\Big(\frac{n \pi r}{L}\Big) ,
\ee
where $L$ is the nonlinear variational parameter of the dimension of length. The second basis, which seems better suited for the problem of interest, is formed by the eigenfunctions of a spherically symmetric harmonic oscillator 
\be
\label{SSH-basis}
\varphi^{\omega}_n(r)=\bigg(\frac{2n! \,\omega^{\frac{3}{2}+l}}{\Gamma \big(n+l+\frac{3}{2}\big)}\bigg)^{\!\! 1/2}
r^{l+1} L_n^{l+1}(\omega r^2) e^{-\frac{\omega x^2}{2}}, 
\ee
where $L_n^{l+1}(\omega r^2)$ are the generalized Laguerre polynomials and $\omega$ is the nonlinear variational parameter of the dimension of inverse length square. Since we are interested in a $s$-wave bound state we put $l=0$ in Eq.~(\ref{SSH-basis}).

Solving the Schr\"odinger equation (\ref{Schroedinger-eq}) with the potential (\ref{potential-1/2}) and with the face values of the parameters given in Table~\ref{table-potential} (in this case the potential is pure real), we have found almost the same binding energy $E_B = - 28.309$~MeV for the trigonometric basis (\ref{trigonometric-basis}) with $L= 7.5$~fm and  $E_B = - 28.310$~MeV for the basis of spherically symmetric harmonic oscillator (\ref{SSH-basis}) with $\omega = 5$~MeV. In both cases the number of basis functions has equaled $N= 90$. Increasing this number has not significantly lowered  the binding energy. The obtained modulus squared of the wave function (\ref{wave-function}) is shown as blue solid line in Fig.~\ref{fig-wave-fun-potential} and it is indistinguishable for both bases of trial functions. 

\begin{figure}[t]
\centering
\includegraphics[width=8.5cm]{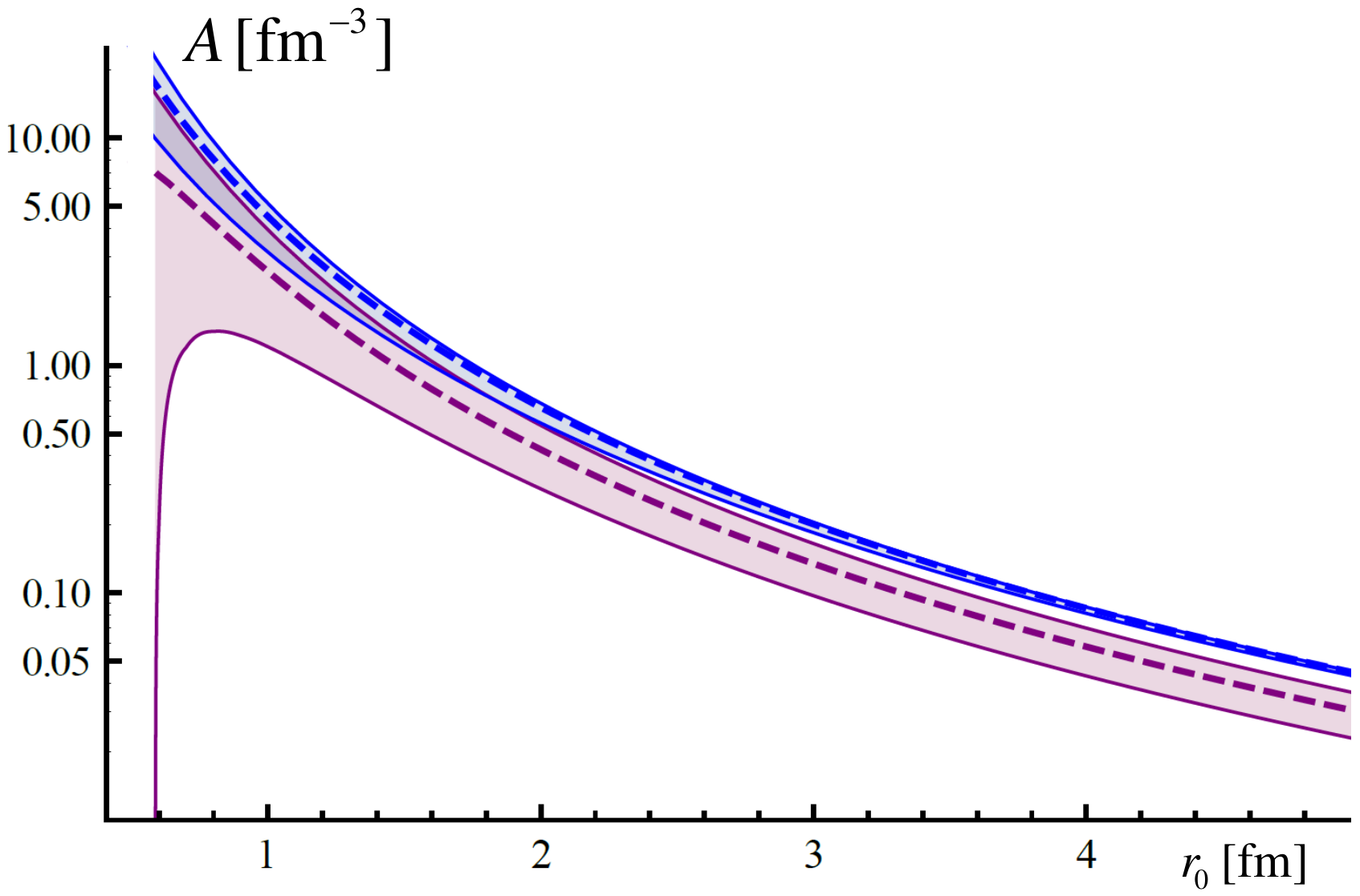}
\vspace{-3mm}
\caption{The bound state formation rate computed directly from Eq.~(\ref{form-rate}) and given by the sum rule (\ref{sum-rule}).}
\label{fig-final-check}
\end{figure}

Looking for the maximal and minimal values of the bound state formation rate (\ref{form-rate}), which enters the sum rule (\ref{sum-rule}), we have solved the Schr\"odinger equation (\ref{Schroedinger-eq}) for various sets of parameter values from the error bands given in Table~\ref{table-potential}. When the parameter $\gamma$ is non-vanishing the potential (\ref{potential-1/2}) acquires an imaginary contribution. 

The variational method, which is described above, can be generalized to a complex-valued potential by using the transformation of complex rotation $U$:
\be
\hat{r} \mapsto U\hat{r}U^{-1} = \hat{r} e^{i\theta}, ~~~~ 
\hat{p}_r \mapsto  U \hat{p}_r U^{-1} = \hat{p}_r e^{-i\theta},
\ee
where $\hat{r}, ~\hat{p}_r$ are the operators of radial position and momentum and $\theta$ is a real parameter and $\theta \in [0,\pi/2]$. The spectrum of the complex-rotated Hamiltonian $\hat{H}_{\theta}=U \hat{H}U^{-1}$, which is no longer Hermitian, is a subject to the Balslev-Combes theorem \cite{Balslev:1971vb,Moiseyev:2011}. The theorem states that the real bound-state and complex eigenvalues are the same as those of the original Hamiltonian $H$. However, the eigenvalues of continuous spectrum of the original Hamiltonian $H$ are rotated down by an angle of $2\theta$ into the lower half-plane of complex energy. Since the eigenvalues of the rotated Hamiltonian are obtained using approximate numerical methods, the eigenvalues, which should be independent of $\theta$, are typically weakly dependent. Therefore, one applies a stabilization procedure looking for such an angle $\theta_{\rm opt}$ that the eigenvalue $E^\theta$ is a stationary solution 
\be
\label{omegaopty}
\frac{d E^{\theta}}{d \theta} \bigg|_{\theta=\theta_{\rm opt}}=0. 
\ee

Using the presented method, we have solved the Schr\"odinger equation (\ref{Schroedinger-eq}) including the maximal imaginary contribution to the potential (\ref{potential-1/2}). The binding energy and wave function acquire imaginary parts. Solving the Schr\"odinger equation (\ref{Schroedinger-eq}) with the face values of real potential parameters and the maximal value of imaginary parameter $\gamma = - 5.4$, we have found the eigenenergy as $E_B = - 28.309 -  0.112 \, i$~MeV. 
We note that the finite value of the parameter $\gamma$ has not changed the real part of the eigenenergy. The modulus square of the wave function found with $\gamma = - 5.4$ (and the optimal value of angle $\theta_{\rm opt} = 0.3$) is shown as the dashed red line in Fig.~\ref{fig-wave-fun-potential}. As one can see, the cases $\gamma = 0$ and $\gamma = -5.4$  are hardly distinguishable. We can therefore conclude that the effects of the admissible value of the imaginary part of the potential are truly small. 

\subsection{Final comparison}

We are now ready to check whether the sum rule (\ref{sum-rule}) is satisfied. The bound state formation rate given by Eq.~(\ref{form-rate}) with the wave function obtained in Sec.~\ref{sec-bound-state} and the Gaussian source function (\ref{Gauss-source-r}) is shown in Fig.~\ref{fig-final-check} as a function of the source size parameter $r_0$. In the same figure we also show the bound state formation rate given by the sum rule (\ref{sum-rule}). The upper limit of the momentum integral $q_{\rm max}$ is big enough that the integral saturates. The blue dashed line represents the formula (\ref{form-rate}) with the face values of the potential parameters. The shaded blue band is obtained by varying the real potential parameters within the range of their maximal errors given in Table~\ref{table-potential}. The magenta dashed line is the formation rate determined by the left-hand-side of Eq.~(\ref{sum-rule}) with the correlation functions computed with the face values of the scattering parameters. The shaded magenta band corresponds to changes of the scattering parameters within their maximal errors given in Table~\ref{table-scatt-parameters}.

As one can see, the blue and red bands overlap that is the sum rule is satisfied for $r_0 \lesssim 2$ fm. In this domain, however, our computation of the correlation functions is not fully reliable, as we use the asymptotic form of the scattering wave function (\ref{scatter}) which is applicable only for sufficiently big inter-particle distances, bigger than the interaction range. Therefore, the fact that the sum rule is satisfied for $r_0 \lesssim 2$ fm is not really meaningful. For $r_0 \gtrsim 2$ fm, where our computation of the correlation functions and bound state formation rate are reliable, the sum rule is not satisfied but the blue and red bands are close to each other.

\section{Discussion and Conclusions}
\label{sec-discussion-conclusion}

Our analysis shows that the correlations functions, which are computed with the scattering parameters of the spin 1/2 and 3/2 channels inferred in the study \cite{Chizzali:2022pjd}, and the bound state formation rate obtained with the bound state wave function, which in turn is determined by the potential also given in the study \cite{Chizzali:2022pjd}, do not satisfy the sum rule (\ref{sum-rule}) but they are not far from it. Presumably, the scattering parameters are not fully consistent with the potential responsible for the bound state. 

\vspace{5mm}

Although the bound state has spin 1/2, not only the correlation function in channel 1/2 but also that in channel 3/2 must be properly tuned for the rule to be satisfied. The correlation function in channel 3/2 is a regulator that removes the divergence but also influences a numerical value of the momentum integral in Eq.~(\ref{sum-rule}). If, for example, the scattering length in channel 3/2 were increased to 6 fm with other parameters unchanged, the sum rule would be perfectly satisfied. Another way to fulfill the sum is probably to explicitly take into account inelastic channels, which, according to the conclusion of the study \cite{Chizzali:2022pjd}, play a negligible role. However, this is a difficult task that we intend to undertake in the future.

\acknowledgements

We are grateful to Francesco Giacosa for fruitful discussions and to Laura Fabbietti, Tetsuo Hatsuda and Yuki Kamiya for useful correspondence. The numerical calculations have been performed using {\it Mathematica}.


\end{document}